\documentclass[runningheads]{svmult}
\usepackage{epsf}   %
\usepackage{psfig}   %
\usepackage{makeidx}   
\usepackage{graphicx}  
\usepackage{subeqnar}  
\usepackage{multicol}  
\usepackage{cropmark} 
\usepackage{physprbb}  
\newcommand{\beq}{\begin{equation}}

\newcommand{\eeq}{\end{equation}}
\newcommand{\bea}{\begin{eqnarray}}
\newcommand{\eea}{\end{eqnarray}}

\newcommand{\vf}{\varphi}

\begin{document}
\title*{Baryonic Q-balls as Dark Matter}
%
%
\toctitle{Q-balls as Dark Matter}
%
%
%
\titlerunning{B-balls as Dark Matter}
%
\author{Alexander Kusenko}
%
%
%
\institute{Department of Physics and Astronomy, UCLA, Los Angeles, CA
90095-1547 \\ 
RBRC, Brookhaven National Laboratory, Upton, New York 11973}

\maketitle              

\begin{abstract}
Supersymmetric extensions of the Standard Model predict the existence of
  Q-balls, some of which can be entirely stable.  Affleck--Dine
  baryogenesis can result in a copious production of stable baryonic
  Q-balls, which can presently exist as a form of dark matter.  
\end{abstract}

\section{Q-balls from Supersymmetry} 

In a class of theories with interacting scalar fields $\phi$ that carry
some conserved global charge, the ground state is a
Q-ball~\cite{q,nts_review}, a lump of coherent scalar condensate that can
be described semiclassically as a non-topological soliton of the form
\begin{equation}
\phi(x,t) = e^{i \omega t} \bar{\phi}(x).
\label{q}
\end{equation}
Q-balls exist whenever the scalar potential satisfies certain conditions
that were first derived for a single scalar degree of freedom~\cite{q} with
some abelian global charge and were later generalized to a theory of many
scalar fields with different charges~\cite{ak_mssm}.  Non-abelian global
symmetries~\cite{nonabelian} and abelian local symmetries~\cite{gauge} can
also yield Q-balls.

For a simple example, let us consider a field theory with a scalar
potential $U(\vf) $  that has a global minimum $U(0)=0$ at $\vf=0$.
Let $U(\vf)$ have an unbroken global\footnote{
Q-balls associated with a local symmetry have been constructed 
\cite{gauge}.  An important qualitative difference is that, in the case of a
local symmetry, there is an upper limit on the charge of a stable Q-ball.} 
U(1) symmetry at the origin, $\vf=0$.  And let the scalar field $\vf$ have
a unit charge with respect to this U(1).

The charge of some field configuration $\vf(x,t)$ is  
\beq
Q= \frac{1}{2i} \int \vf^* \stackrel{\leftrightarrow}{\partial}_t  
\vf \, d^3x . 
\label{Qt}
\eeq
Since a  trivial configuration $\vf(x)\equiv 0$ has zero charge, the
solution that minimizes the energy, 
\beq
E=\int d^3x \ \left [ \frac{1}{2} |\dot{\vf}|^2+
\frac{1}{2} |\nabla \vf|^2 
+U(\vf) \right], 
\label{e}
\eeq
and has a given charge $Q>0$, must differ from zero in some (finite)
domain.  This is a Q-ball.   It is a time-dependent solution, more
precisely it has a time-dependent phase. However, all physical quantities
are time-independent.  Of course, we have not proven that such a 
``lump'' is finite, or that it has a lesser energy than the collection of
free particles with the same charge; neither is true for a general
potential.  A finite-size  Q-ball is a minimum of energy and is
stable with respect to decay into free $\vf$-particles if 
\beq
U(\vf) \left/ \vf^2 \right. = {\rm min},
\ \ {\rm for} \ 
\vf=\vf_0>0 .
\label{condmin}
\eeq

One can show that the equations of motion for a Q-ball in 3+1 dimensions
are equivalent to those for the bounce associated with tunneling 
in 3 Euclidean dimensions in an effective potential $\hat{U}_\omega
(\vf)= U(\vf) - (1/2) \omega^2 \vf^2$, where $\omega$ is such that it
extremizes~\cite{ak_qb}
\beq
{\cal E}_\omega = S_3(\omega) +\omega Q. 
\label{Ew}
\eeq
Here $S_3(\omega)$ is the three-dimensional Euclidean action of the bounce
in the potential $\hat{U}_\omega (\vf)$.  The Q-ball
solution has the form (\ref{q}), 
%
where $\bar{\vf}(x)$ is the bounce. 

The analogy with tunneling clarifies the meaning of condition 
(\ref{condmin}), which simply requires that there exist a value of 
$\omega$, for which $\hat{U}_\omega (\vf)$ is negative for some value of 
$\vf=\vf_0 \neq 0$ separated from the false vacuum by a barrier. 
This condition ensures the  existence of a bounce.  (Clearly, the
bounce does not exist if $\hat{U}_\omega (\vf) \ge 0$ for all $\vf$ because
there is nowhere to tunnel.)  

In the true vacuum, there is a minimal value $\omega_0$, so that only for
$\omega>\omega_0$, $\hat{U}_\omega (\vf)$ is somewhere negative.  If one
considers a Q-ball in a metastable false vacuum, then $\omega_0=0$.  The
mass of the $\vf$ particle is the upper bound on $\omega$ in either
case. Large values of $\omega$ correspond to small charges~\cite{ak_qb}.
As $Q \rightarrow \infty$, $\omega \rightarrow \omega_0$.  In this case,
the effective potential $\hat{U}_\omega (\vf)$ has two nearly-degenerate
minima; and one can apply the thin-wall approximation to calculate the
Q-ball energy~\cite{q}.  For smaller charges, the thin-wall approximation
breaks down, and one has to resort to other methods~\cite{ak_qb}.

The above discussion can be generalized to the case of several fields, 
$\vf_k$, with different charges, $q_k$~\cite{ak_mssm}.  Then the Q-ball is
a solution of the form 
\beq
\vf_k(x,t) = e^{iq_k \omega t} \vf_k(x),
\label{tsol}
\eeq
where $\vf(x)$ is again a three-dimensional bounce associated with
tunneling in the potential 
\beq
\hat{U}_\omega (\vf) = U(\vf)\ - \ \frac{1}{2} \omega^2 \, 
\sum_k q_k^2 \, |\vf_k|^2. 
\label{Uhat}
\eeq
As before, the value of $\omega$ is found by minimizing ${\cal E}_\omega$
in equation  (\ref{Ew}).  The bounce, and, therefore, the Q-ball, exists if 
\begin{eqnarray}
\mu^2 & = & 
2 U(\vf) \left/ \left (\sum_k q_k \vf_{k,0}^2 \right ) \right. = {\rm min},
\ \nonumber \\ & & {\rm for} \ |\vec{\vf}_0|^2 > 0.
\label{condmin1}
\end{eqnarray}

The soliton mass can be calculated by extremizing ${\cal E}_\omega$ in
equation (\ref{Ew}).  If $ |\vec{\vf}_0|^2 $ defined by equation
(\ref{condmin1}) is finite, then the mass of a soliton $M(Q)$ is
proportional to the first power of $Q$: 
\beq
M(Q) = \tilde{\mu} Q, \ \ {\rm if} \ |\vec{\vf}_0|^2 \neq \infty. 
\label{MQ}
\eeq In particular, if $Q\rightarrow \infty$, $\tilde{\mu}\rightarrow \mu$
(thin-wall limit)~\cite{q,nts_review}.  For smaller values of $Q$,
$\tilde{\mu}$ was computed in~\cite{ak_qb}.  In any case, $\tilde{\mu}$ is
less than the mass of the $\phi$ particle by definition (\ref{condmin1}).

However, if the scalar potential grows slower than the second power of
$\phi$, then $|\vec{\vf}_0|^2 = \infty$, and the Q-ball never reaches the
thin-wall regime, even if Q is large.  The value of $\phi$ inside the
soliton extends as far as the gradient terms allow, and the mass of a
Q-ball is proportional to $Q^{p}$, $p<1$.  In particular, if the scalar
potential has a flat plateau $U(\phi) \sim m $ at large $\phi$, then the
mass of a Q-ball is~\cite{dks} 
\beq
M(Q) \sim m Q^{3/4}.
\label{MQflat}
\eeq
This is the case for the stable baryonic Q-balls in the MSSM discussed
below. 

It turns out that all phenomenologically viable supersymmetric extensions
of the Standard Model predict the existence of non-topological
solitons~\cite{ak_mssm} associated with the conservation of baryon and
lepton number. If the physics beyond the standard model reveals some
additional global symmetries, this will further enrich the spectrum of
Q-balls~\cite{Demir}.  The MSSM admits a large number of different Q-balls,
characterized by (i) the quantum numbers of the fields that form a
spatially-inhomogeneous ground state and (ii) the net global charge of this
state.

First, there is a class of Q-balls associated with the tri-linear
interactions that are inevitably present in the MSSM~\cite{ak_mssm}.  The
masses of such Q-balls grow linearly with their global charge, which can be
an arbitrary integer number~\cite{ak_qb}.  Baryonic and leptonic Q-balls of
this variety are, in general, unstable with respect to their decay into
fermions.  However, they could form in the early universe through the
accretion of global charge~\cite{gk,ak_pt} or, possibly, in a first-order
phase transition~\cite{s_gen}.

The second class~\cite{dks} of solitons comprises the Q-balls whose VEVs
are aligned with some flat directions of the MSSM.  The scalar field inside
such a Q-ball is a gauge-singlet~\cite{kst} combination of squarks and
sleptons with a non-zero baryon or lepton number.  The potential along a
flat direction is lifted by some soft supersymmetry-breaking terms that
originate in a ``hidden sector'' of the theory at some scale $\Lambda_{_S}$
and are communicated to the observable sector by some interaction with a
coupling $g$, so that $g \Lambda \sim 100$~GeV.  Depending on the strength
of the mediating interaction, the scale $\Lambda_{_S}$ can be as low as a
few TeV (as in the case of gauge-mediated SUSY breaking), or it can be some
intermediate scale if the mediating interaction is weaker (for instance,
$g\sim \Lambda_{_S}/m_{_{Planck}}$ and $\Lambda_{_S}\sim 10^{10}$~GeV in
the case of gravity-mediated SUSY breaking).  For the lack of a definitive
scenario, one can regard $\Lambda_{_S}$ as a free parameter.  Below
$\Lambda_{_S}$ the mass terms are generated for all the scalar degrees of
freedom, including those that parameterize the flat direction.  At the
energy scales larger than $\Lambda_{_S}$, the mass terms turn off and the
potential is ``flat'' up to some logarithmic corrections.  If the Q-ball
VEV extends beyond $\Lambda_{_S}$, the mass of a soliton~\cite{dks,ks} is
no longer proportional to its global charge $Q$, but rather to $Q^{3/4}$.
A hybrid of the two types is yet another possibility~\cite{hybrid}. 

This allows for the existence of some entirely stable Q-balls with a large
baryon number $B$ (B-balls).  Indeed, if the mass of a B-ball is $M_{_B} \sim
({\rm 1~TeV}) \times B^{3/4}$, then the energy per baryon number
$(M_{_B}/B)\sim ({\rm 1~TeV}) \times B^{-1/4}$ is less than 1~GeV for $B >
10^{12}$.  Such large B-balls cannot  dissociate into protons and neutrons
and are entirely stable thanks to the conservation of energy and the baryon
number.  If they were  produced in the early universe, they would exist at
present as a form of dark matter~\cite{ks}.

\section{Fragmentation of Affleck--Dine Condensate into Q-balls}
Several mechanisms could lead to formation of B-balls and L-balls in the
early universe. First, they can be produced in the course of a phase
transition~\cite{s_gen}.  Second, thermal fluctuations of a baryonic and
leptonic charge can, under some conditions, form a Q-ball.  Finally, a
process of a gradual charge accretion, similar to nucleosynthesis, can take
place~\cite{gk,ak_pt,dew}.  However, it seems that the only process that can
lead to a copious production of very large, and, hence, stable, B-balls, is
fragmentation of the Affleck-Dine condensate~\cite{ks}. 

At the end of inflation, the scalar fields of the MSSM develop some large
expectation values along the flat directions, some of which have a non-zero
baryon number~\cite{ad}. Initially, the scalar condensate has the form
given in eq.~(\ref{q}) with $\bar{\phi}(x)= const$ over the length scales
greater than a horizon size. One can think of it as a universe filled with
Q-matter.  The relaxation of this condensate to the potential minimum is
the basis of the Affleck--Dine (AD) scenario for baryogenesis.

It was often assumed that the condensate remains spatially homogeneous from
the time of formation until its decay into the matter baryons.  This
assumption is, in general, incorrect.  In fact, the initially homogeneous
condensate can become unstable~\cite{ks} and break up into Q-balls whose
size is determined by the potential and the rate of expansion of the
Universe.  B-balls with $12 < \log_{10} B < 30$ can form naturally
from the breakdown of the AD condensate.  These are entirely
stable if the flat direction is ``sufficiently flat'', that is if the
potential grows slower than $\phi^2$ on the scales or the order of
$\bar{\phi}(0)$.   The evolution of the primordial condensate can be
summarized as follows: 

\vspace{3mm}
\psfig{figure=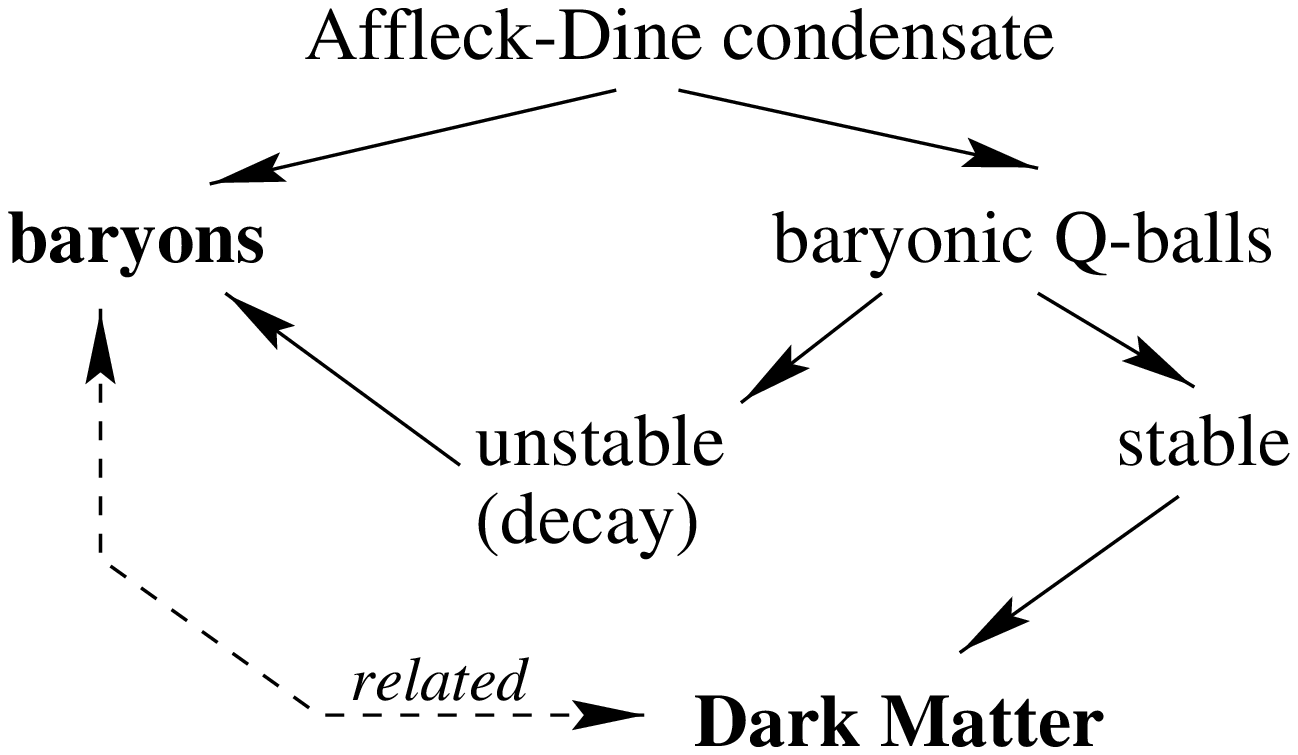,height=2.1in,width=4.5in}

This process has been analyzed analytically~\cite{ks,em} in the linear
approximation.  Recently, some impressive numerical simulations of Q-ball
formation have been performed~\cite{kasuya}; they confirm that the
fragmentation of the condensate into Q-balls occurs in some Affleck-Dine
models.  The global charges of Q-balls that form this way are model
dependent.  The subsequent collisions~\cite{ks,Axenides:1999hs} can further
modify the distribution of soliton sizes.

\begin{figure}
\setlength{\epsfxsize}{3.4in}
\centerline{\epsfbox{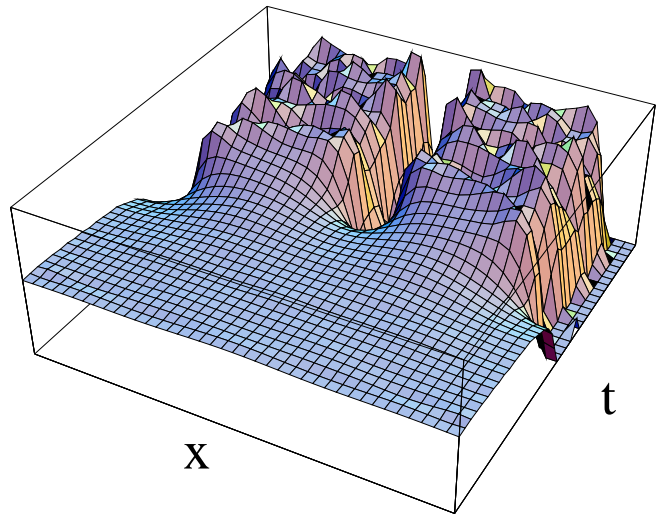}}
\caption{ 
The charge density per comoving volume in (1+1) dimensions for a sample
potential analyzed numerically during the fragmentation of the condensate
into Q-balls. 
}
\label{fig_charge}
\end{figure}

In supersymmetric extensions of the Standard Model, Q-ball formation occurs
along flat directions of a certain type, which appear to be generic in the
MSSM~\cite{Enqvist:2000gq}.

\section{SUSY Q-balls as Dark Matter} 

Conceivably, the cold dark matter in the Universe can be made up entirely
of SUSY Q-balls.  Since the baryonic matter and the dark matter share the
same origin in this scenario, their contributions to the mass density of
the Universe are related.  Most of dark-matter scenarios offer no
explanation as to why the
observations find $\Omega_{_{DARK}} \sim \Omega_{B} $ within an order of
magnitude.  This fact is extremely difficult to explain in models that
invoke a dark-matter candidate whose present-day abundance is determined by
the process of freeze-out, independent of baryogenesis.  
If one doesn't want to accept this equality
as fortuitous, one is forced to hypothesize some {\it ad hoc}
symmetries~\cite{kaplan} that could relate the two quantities.  In the MSSM
with AD baryogenesis, the amounts of dark-matter Q-balls and the ordinary
matter baryons are related~\cite{ks}; one
predicts~\cite{lsh} $\Omega_{_{DARK}} = \Omega_{B} $ for
B-balls with $B \sim 10^{26}$. However, the size of Q-balls depends on the
supersymmetry breaking terms that lift the flat direction.  
The required size is in the middle of the range of
Q-ball sizes that can form in the Affleck--Dine
scenario~\cite{ks,em,kasuya}.  Diffusion effects may force the Q-balls
sizes to be somewhat smaller, $B\sim 10^{22} -  10^{24}$, if they are to be
CDM and to generate the baryon asymmetry of the universe through partial 
evaporation~\cite{Banerjee:2000mb}.

\begin{figure}
\setlength{\epsfxsize}{4.5in}
\epsfclipon
\centerline{\epsfbox{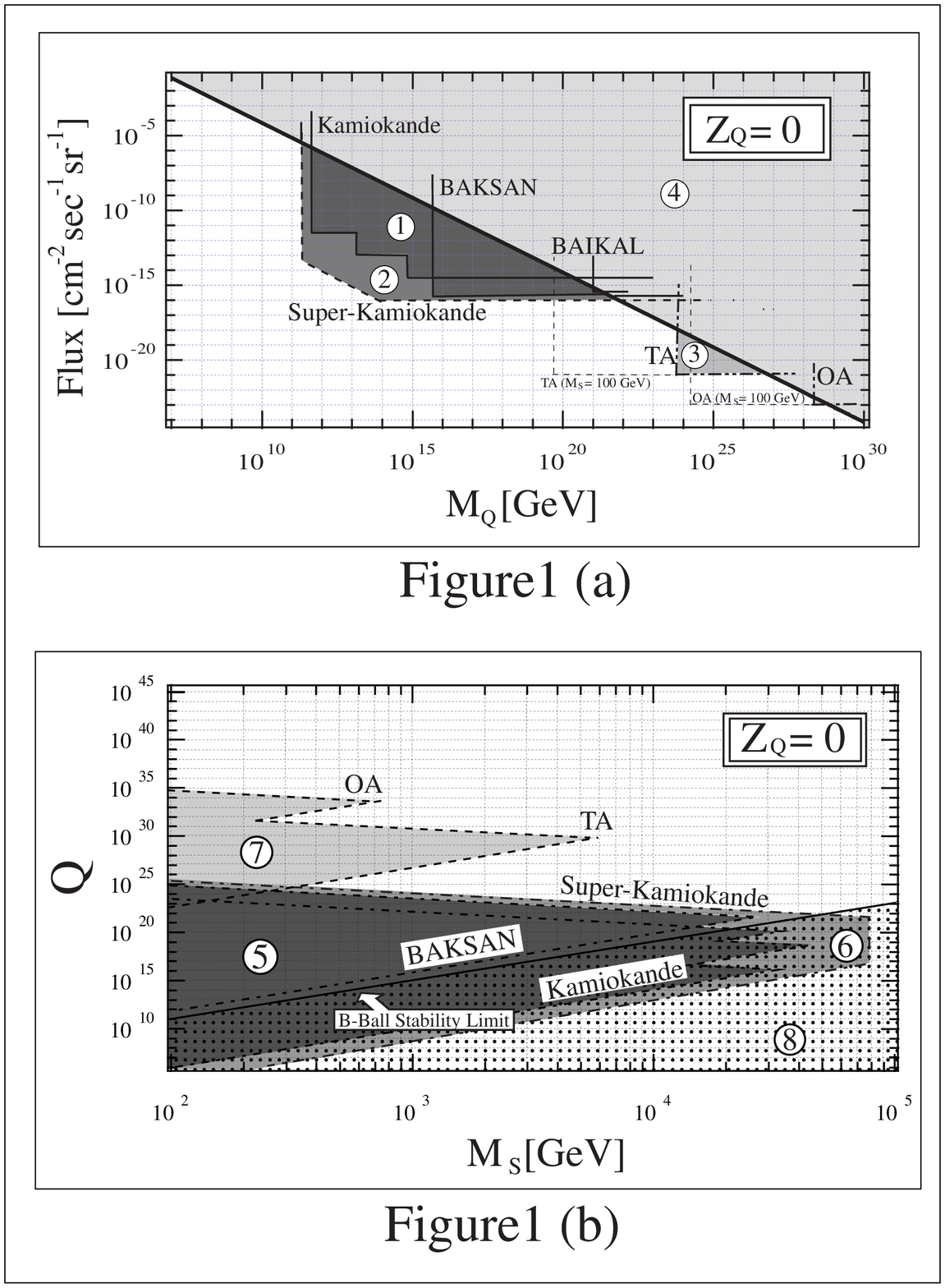}}
\caption{ The resent limits on the baryon numbers of electrically neural
dark-matter Q-balls from a paper by J.~Arafune {\em et al.}~\cite{arafune}.
}
\epsfclipoff
\label{fig_limits}
\end{figure}

The value $B\sim 10^{26}$ is well within the present experimental limits 
on the baryon number of an average relic B-ball, under the assumption
that all or most of cold dark matter is made up of Q-balls.  On their 
passage through matter, the electrically neutral baryonic SUSY Q-balls can
cause a proton decay, while the electrically charged B-balls produce 
massive ionization.  Although the condensate inside a Q-ball is
electrically neutral~\cite{kst}, it may pick up some electric charge
through its interaction with matter~\cite{kkst}.  Regardless of its ability
to retain electric charge, the Q-ball would produce a straight track in a
detector and would release the energy of, roughly, 10 GeV/mm.  The present
limits~\cite{kkst,exp,arafune} constrain the baryon number of a relic
dark-matter B-ball to be greater than $10^{22}$.  Future experiments are
expected to improve these limits.  It would take a detector with the area of
several square kilometers to cover the entire interesting range $B\sim
10^{22} ... 10^{30}$.

\section{Star Wreck: the Q-ball Invasion}

In non-supersymmetric theories, nuclear matter of neutron stars is the
lowest-energy state with a given baryon number\footnote{I remind the reader
  that black holes do not have a well-defined baryon number.}.  In 
supersymmetric theories, however, a Q-ball with baryon number $10^{57}$ can
be lighter than a neutron star.  I am going to describe a process that 
can transform a neutron star into a very large B-ball.  The time scale
involved is naturally of the order of billion years.

Dark-matter superballs pass through the ordinary stars and planets with a
negligible change in their velocity.  However, both SECS and SENS stop in
the neutron stars and accumulate there~\cite{sw}.  As soon as the first
Q-ball is captured by a neutron star, it sinks to the center and begins to
absorb the baryons into the condensate.  High baryon density inside a
neutron star makes this absorption very efficient, and the B-ball grows at
the rate that increases with time due to the gradual increase in the
surface area.  After some time, the additional dark-matter Q-balls that
fall onto the neutron star make only a negligible contribution to the
growth of the central Q-ball~\cite{sw}.  So, the fate of the neutron star
is sealed when it captures the first superball.

According to the discussion in section 3, the energy per unit baryon number
inside the relic B-ball is less than that in nuclear matter.   Therefore,
the absorption process is accompanied by the emission of heat carried away  
by neutrinos and photons.   We estimate that this heating is too weak to
lead to any observable consequences.  However, the absorption of nuclear
matter by a baryonic Q-ball causes a gradual decrease in the mass of the
neutron star.  

Neutron stars are stable in some range of masses.  In particular, there is
a minimal mass (about 0.18 solar mass), below which the force of gravity is
not strong enough to prevent the neutrons from decaying into protons and
electrons.  While the star is being consumed by a superball, its mass
gradually decreases, reaching the critical value eventually.  At that
point, a mini-supernova explosion occurs~\cite{minisn}, which can be
observable.  Perhaps, the observed gamma-ray bursts may originate from an
event of this type.  A small geometrical size of a neutron star and a large
energy release may help reconcile the brightness of the gamma-ray bursts
with their short duration. 

Depending on the MSSM parameters, the lifetime of a neutron star $t_s$ can
range from 0.01 Gyr to more than 10 Gyr~\cite{sw}:
\begin{equation}
t_s \sim \frac{1}{\beta} \times 
\left ( \frac{m}{200\, {\rm GeV}} \right )^{5} {\rm Gyr}, 
\end{equation}
where $\beta$ is some model-dependent quantity expected to be of order
one~\cite{sw}.   The ages of pulsars set the limit $t_s> 0.1$~Gyr.   

It is interesting to note that $t_s$ depends on the fifth power of the
mass parameter $m$ associated with supersymmetry breaking.  If the 
mini-supernovae are observed (or if the connection with gamma-ray bursts is
firmly established), one can set strict constraints on the supersymmetry
breaking sector from the rate of neutron star explosions. 

The naturally long time scale is intriguing.

\section{B-ball Baryogenesis}

An interesting scenario that relates the amounts of baryonic and dark matter
in the Universe, and in which the dark-matter particles are produced from
the decay of unstable B-balls was proposed by Enqvist and
McDonald~\cite{em}.  

\section{Phase Transitions Precipitated by Solitosynthesis} 

In the false vacuum, a rapid growth of non-topological
solitons~\cite{gk} can precipitate an otherwise impossible or slow phase
transition~\cite{ak_pt}.  

\begin{figure}
\setlength{\epsfysize}{2.7in}
\centerline{\epsfbox{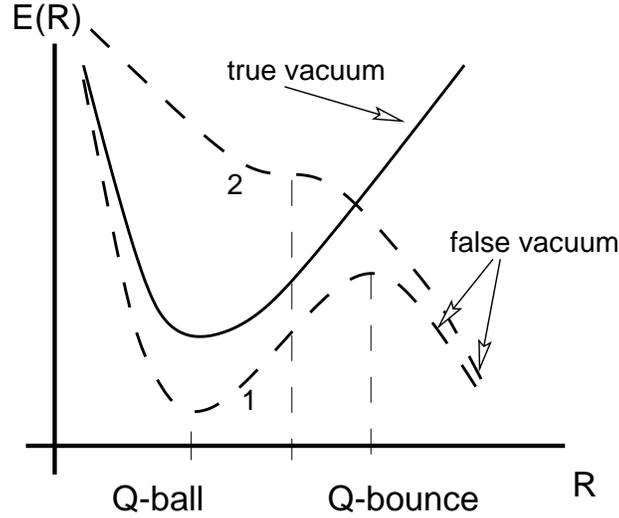}}
\caption{
Energy (mass) of a soliton as a function of its size.  In the true vacuum, 
Q-ball is the global minimum of energy (solid curve).  In the false vacuum,
if the charge is less than some critical value, there are two solutions: a
``stable'' Q-ball, and an unstable ``Q-bounce'' (dashed curve 1) .  In the
case of a critical charge (curve 2), there is only one solution, which is
unstable. 
} 
\label{fig2}
\end{figure}

Let us suppose the system is in a metastable false vacuum that preserves
some U(1) symmetry.  The potential energy in the Q-ball interior is
positive in the case of a true vacuum, but negative if the system is in the
metastable false vacuum. In either case, it grows as the third power of the
Q-ball radius $R$.  The positive contribution of the time derivative to the
soliton mass can be written as $Q^2/\int \bar{\phi}^2(x)d^3x \propto
R^{-3}$, and the gradient surface energy scales as $R^2$.  In the true
vacuum, all three contributions are positive and the Q-ball is the absolute
minimum of energy (Fig.~\ref{fig2}).  However, in the false vacuum, the
potential energy inside the Q-ball is negative and goes as $\propto -R^3$.
As shown in Fig.~\ref{fig2}, for small charge $Q$, there are two stationary
points, the minimum and the maximum.  The former corresponds to a Q-ball
(which is, roughly, as stable as the false vacuum is), while the latter is
a critical bubble of the true vacuum with a non-zero charge.

There is a critical value of charge $Q=Q_c$, for which the only stationary
point is unstable.  If formed, such an unstable bubble will expand.

If the Q-ball charge increases gradually, it eventually reaches the
 critical value.  At that point Q-ball expands and converts space into a
 true-vacuum phase.  In the case of tunneling, the critical bubble is
 formed through coincidental coalescence of random quanta into an extended
 coherent object.  This is a small-probability event.  If, however, a
 Q-ball grows through charge accretion, it reaches the critical size with
 probability one, as long as the conditions for growth~\cite{ak_pt} are
 satisfied.  The phase transition can proceed at a much faster rate than it
 would by tunneling.

\section{Conclusion}

Supersymmetric models of physics beyond the weak scale offer two plausible 
candidates for cold dark matter.  One is the lightest supersymmetric
particle, which is stable because of R-parity.  Another one is a
stable non-topological soliton, or Q-ball, carrying some baryonic charge. 

SUSY Q-balls make an appealing dark-matter candidate because their
formation is a natural outcome of Affleck--Dine baryogenesis and requires
no unusual assumptions. 

In addition, formation and decay of unstable Q-balls can have a dramatic
effect on baryogenesis, dark matter, and the cosmic microwave
background.  Production of unstable Q-balls in the false vacuum can cause an
unusually fast first-order phase transition.

My work was supported in part by the US Department of Energy grant
DE-FG03-91ER40662, Task C, and by a UCLA Council on Research faculty  grant.

%

\end{document}